\documentclass{wsk}

\newcommand{\gsim}{\lower.7ex\hbox{$\;\stackrel{\textstyle>}{\sim}\;$}}
\newcommand{\lsim}{\lower.7ex\hbox{$\;\stackrel{\textstyle<}{\sim}\;$}}

\begin{document}

\title{Flavor Violation as a Probe of the \\ MSSM Higgs Sector}

\author{Christopher Kolda}

\address{Theoretical Physics Group, Lawrence Berkeley National Laboratory
  \\ University of California, Berkeley, CA 94720, USA}


\maketitle

\abstracts{At moderate to large $\tan\beta$, it is no longer possible
  to simultaneously diagonalize the masses of quarks and their
  couplings to the neutral Higgs bosons. The resulting flavor
  violations of the form $\bar b_Rd_L\phi$ and $\bar b_Rs_L\phi$ do
  not generate large meson--anti-meson mixing amplitudes but do
  generate large contributes to rare decays such as
  $B_s\to\mu\mu$. Run~II of the Tevatron willl probe a large region of 
  interesting MSSM parameter space through this decay channel. This 
  talk is based on results obtained with K.S.~Babu and presented in [1].}

One of the most interesting facts about the minimal supersymmetric
standard model (MSSM) is that it is {\em not}\/ the
supersymmetric minimal standard model. 
Holomorphy and anomaly cancellation force 
an additional Higgs doublet onto the model. 
Yet despite this extra Higgs doublet, it is still tempting to think 
of the minimal standard model as being the low-energy limit of the MSSM
rather than its 2-Higgs doublet extension (the 2HDSM).

In a series of papers~\cite{bk,bkplus}
my coauthors and I have demonstrated
that there are important new classes of phenomena
associated with the 2HDSM-like limit of the MSSM which do not occur
in a standard model-like limit. (Similar effects have been studied in [3].)
These phenomena arise 
because the neutral Higgs bosons of the MSSM need not couple
to fermions with strength proportional to the fermion mass, contrary
to common folklore. 

Consider first the non-supersymmetric 2HDSM. 
It is well known that such models must be carefully 
arranged in order to avoid large flavor-changing neutral currents (FCNCs).
In particular, one must ensure that the couplings of the neutral Higgs 
states to fermions are proportional to the fermion masses. Were they not,
Yukawa couplings would not be diagonal in the fermion mass eigenbases.
The most economical method for guaranteeing this proportionality is
to segregate the two Higgs doublets, allowing one to couple only to
the $\bar U_RQ_L$ bilinear and the other to couple only to $\bar D_R
Q_L$:
\begin{equation}
-{\cal L}=\bar U_R {\bf Y_U} Q_L H_u + \bar D_R {\bf Y_D} Q_L H_d + h.c.
\label{2hdsm}
\end{equation}
where $U,D,Q$ are 3-vectors in flavor space and ${\bf Y_{U,D}}$ are
$3\times3$ matrices. Thus $H_u$ is alone responsible for giving mass
to $u$-quarks and $H_d$ to $d$-quarks. When the Higgs fields mix 
to form a Higgs mass eigenstate $\phi=\{h^0,H^0,A^0\}$, 
the coupling to any given fermion will 
either come through the $H_u$ or the $H_d$ component of $\phi$, but
not both. Therefore $\phi$ also couples proportional to fermion
mass, though with reduced strength.

This type of model, often called a ``Type-II'' model, is also
stable against radiative corrections. 
Undesirable couplings of the form 
\begin{equation}
-{\cal L}_{{}_{bad}}=\bar U_R {\bf Y_U'} Q_L H_d^* 
+\bar D_R {\bf Y_D'} Q_L H_u^* + h.c.
\label{danger}
\end{equation}
(which would generate FCNCs) 
are forbidden by a $Z_2$ symmetry under
which one of the two Higgs doublets is odd.

The MSSM is also a type-II model, but it possesses no such $Z_2$
symmetry. The superpotential of the MSSM is given by:
\begin{equation}
W=\hat U_R {\bf Y_U} \hat Q_L \hat H_u + \hat D_R {\bf Y_D} \hat Q_L
\hat H_d + \mu \hat H_u \hat H_d
\label{W}
\end{equation}
where I am ignoring leptons. It is the presence of 
a non-zero $\mu$-term which breaks the $Z_2$; and 
experimentally we know $\mu\neq0$ due to the absence of a massless
neutralinos in our colliders. However, the MSSM does not need any
$Z_2$ to preserve the form of the Yukawa interactions. As long as SUSY is
unbroken, holomorphy and the non-renormalization theorem guarantee
that the dangerous Yukawa interactions of Eq.~(\ref{danger}) cannot arise.

But SUSY {\em is}\/ broken. And without any $Z_2$ to protect the form of
(\ref{2hdsm}), there is nothing to prevent couplings such as those in
(\ref{danger}) from arising. In fact, they are known to arise after 
SUSY-breaking. In its
present form, this observation was first made by Hall, Rattazzi and
Sarid\cite{hrs} (HRS). They found that $d$-quarks can receive a 1-loop
correction to their masses from the operator $\bar D_R Q_L
H_u^*$. Though the coefficient of this new operator is small ($\sim
\frac{1}{16\pi^2}$), the contribution of this operator to the $d$-quarks
masses is enhanced by the ratio of the vev of $H_u$ to $H_d$, that
is, by $\tan\beta$. Further, the leading diagrams do not decouple in
the heavy SUSY mass limit, so long as all SUSY mass scales became
large together.

The presence of a $\bar D_R Q_L H_u^*$ operator should be enough clue
that something interesting will happen if we leave the full flavor
structure of the HRS calculation intact.
HRS considered two diagrams (shown in Fig.~1) which contribute to the
$d$-quark masses, the larger mediated by gluinos, the smaller by
charged Higgsinos. 
\begin{figure}[t]
\centering
\epsfysize=0.9truein
\epsfbox{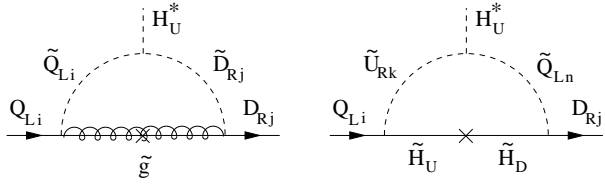} 
\caption{Diagrams which contribute to $\bar D_R Q_L H_u^*$ operator.}
\end{figure}
Both diagrams generate couplings between $\bar D_R
Q_L$ and $H_u^*$. The gluino diagram, however, has a trivial flavor
structure. Thus when we
diagonalize the canonical $d$-quark mass term, we will simultaneously
diagonalize this new contribution, preventing any
flavor-violation. The second diagram is not so trivial thanks to the
top squarks in the loop. Here the flavor structure of the new operator
is given by ${\bf Y_DY_U}^\dagger {\bf Y_U}$, which is not diagonalized in the
same basis as ${\bf Y_D}$. Thus it is this second diagram which generates
the flavor-changing. Simplify by diagonalizing the $u$-quark
Yukawa interactions (similar effects are not sizable in the
$u$-sector), leaving an effective Lagrangian in the $d$-sector:
\begin{equation}
-{\cal L}=\bar D_R\, {\bf DV^0}^\dagger Q_L H_d + \bar D_R\, {\bf
DV^0}^\dagger\left[
\epsilon_g + \epsilon_u{\bf U}^\dagger{\bf U}\right] Q_L H_u^* + h.c.
\end{equation}
where ${\bf U},{\bf D}$ are the diagonalized Yukawa matrices; $\epsilon_g$ and
 $\epsilon_u$ are the contributions from the two diagrams in Fig.~1:
\begin{equation}
\epsilon_g \simeq (2\alpha_3/3\pi) \mu^* M_3, \quad\quad
\epsilon_u \simeq (1/16\pi^2) \mu^* A_U,
\end{equation}
and $V^0$ is a unitary matrix which becomes
the CKM matrix $V$ as $\epsilon_u\to0$. 

Now consider only the neutral current contributions.
In the $u$-quark mass eigenbasis, the SU(2)
partner of $U_L$ is not $D_L$ but rather $D_L'=VD_L$. Then the
neutral current Lagrangian has the form:
\begin{equation}
-{\cal L}_{{}_{N\!C}}
=\bar D_R\, {\bf DV^0}^\dagger{\bf V} D_L H_d^0 + \bar D_R\, {\bf
DV^0}^\dagger\left[
\epsilon_g + \epsilon_u{\bf U}^\dagger{\bf U}\right]{\bf V} D_L H_u^{0*} + h.c.
\end{equation}
The matrix $V^0$ can be calculated explicitly since it must
diagonalize the $d$-quark mass matrix:
$V^{0\dagger} {\cal Y}^\dagger {\cal Y} V^0
 = \mbox{diag}\,(m_d^2,m_s^2,m_b^2)$
where
\begin{equation}
{\cal Y}={\bf DV^0}^\dagger\left[1+\tan\beta\left(\epsilon_g+\epsilon_u
{\bf U}^\dagger{\bf U}\right)\right].
\end{equation}
Keeping only the Yukawa couplings of the third generation, one can do
the algebra, pulling out the flavor-changing pieces:
\def\chifc{\chi_{\!{}_{F\!C}}}
\begin{equation}
{\cal L}_{{}_{F\!C\!N\!C}}
=\frac{\bar y_b V^*_{tb}}{\sin\beta}\chifc\left[V_{td}\bar b_R
d_L + V_{ts} \bar b_R s_L\right]\left(\cos\beta H_u^{0*}-\sin\beta
H_d^0\right) + h.c.
\end{equation}
where $\bar y_b=m_b/v_d$ ($\simeq1$ for large $\tan\beta$) and 
$\chifc\simeq -\epsilon_u y_t^2\tan\beta$.
Note that $\chifc$ is proportional to $\epsilon_u$ as expected
(complete expression in [1]).
Of course, the $H_u$ and $H_d$ are not mass eigenstates, but it is
simple to go to their mass eigenbasis\cite{bk} so we will not show it
here. Suffice to say that the flavor-changing coupling of the $h^0$ (lightest
Higgs) goes to zero for large $m_{A^0}$, while those for the other Higgs
bosons do not.

Where should we look for signs of this flavor changing? For most
sources of FCNCs, it is meson--anti-meson mixing which provides the
best opportunity for testing new effects. However here this is not the
case. One can show that there is a remnant flavor symmetry forbidding
(at lowest order) $\Delta B=2$ operators such as
$B^0-\bar B^0$ mixing. This left-over symmetry is present in the MSSM
only (not in generic two-Higgs models) 
and only at lowest order. There are $\Delta B=2$ contributions at higher order,
but these have an additional loop suppression and are
therefore highly suppressed. Thus, it turns out that meson--anti-meson
mixing is a poor probe of this form of flavor-changing. 

Rare $B$-decays ($\Delta B=1$) provide another route, one which is not
suppressed by residual flavor symmetries. Since the mediation is via a
Higgs, final states are preferred according to their mass. Thus
$B\to\tau\tau$ would be ideal were it not so difficult
experimentally. On the other hand, $B\to ee$ or $B\to X\nu\nu$ are
highly suppressed. The optimal case is $B\to\mu\mu$ which was studied in
[1]. (Right behind is $B\to X\mu\mu$, only slightly weaker than the
purely leptonic case.) This decay can be probed at the
Tevatron where $B_d$ and $B_s$ are produced in a ratio of about
3:1. However, the relative widths for $B_{d,s}\to\mu\mu$ scale as
$(V_{td}/V_{ts})^2\sim1/25$, so it is $B_s\to\mu\mu$ which will
interest us.

For lack of space, only the final results will be shown here. Current
CDF limits\cite{cdf} of $Br(B_s\to\mu\mu)<2\times10^{-6}$ 
constrain $m_{A^0}\lsim200\,$GeV given very large values of
$\tan\beta\simeq 40-70$ and
degenerate SUSY masses. At Run~II, that bound will increase to about
400 GeV after one year (1 fb${}^{-1}$) and 650 GeV after several years
(5 fb${}^{-1}$). As long as $m_{A^0}\lsim 1\,$TeV and $\tan\beta$ is
large, the MSSM contributions to $B\to\mu\mu$ will exceed the SM
prediction and be observed before or at the LHC. And because these
contributions persist even for very heavy SUSY masses, the decay $B\to\mu\mu$
may be our first hint of SUSY, long before SUSY partners are directly
produced and observed.

\section*{Acknowledgments}
This work was supported in part by the U.S.\ Department of Energy under
contract DE-AC03-76SF00098.


\begin{thebibliography}{99}
\bibitem{bk} K.S.~Babu and C.~Kolda, 
\Journal{\PRL}{84}{228}{2000}. 

\bibitem{bkplus} K.S.~Babu, C.~Kolda, J.~March-Russell and F.~Wilczek, 
\Journal{\PRD}{59}{016004}{1998}; 
K.S.~Babu and C.~Kolda,
\Journal{\PLB}{451}{77}{1999}. 

\bibitem{others} C.~Hamzaoui, M.~Pospelov and M.~Toharia,
\Journal{\PRD}{59}{095005}{1999}; 
M.~Carena, S.~Mrenna and C.~Wagner, 
\Journal{\PRD}{60}{075010}{1999}; 
F.~Borzumati, G.~Farrar, N.~Polonsky and S.~Thomas,
\Journal{\NPB}{555}{53}{1999}. 

\bibitem{hrs} L.~Hall, R.~Rattazzi and U.~Sarid,
\Journal{\PRD}{50}{7048}{1994}. 

\bibitem{cdf} F.~Abe {\it et al} [CDF Collaboration],
\Journal{\PRD}{57}{3811}{1998}. 

\end{thebibliography}
\end{document}